\documentclass[letterpaper, 10pt, conference]{IEEEconf}%
\usepackage{amssymb}
\usepackage{amsmath}
\usepackage{amsfonts}
\usepackage{graphicx}
\usepackage{multirow}
\usepackage{epsfig}
\usepackage[belowskip=0pt,aboveskip=10pt]{caption}%
\begin{document}

\title{{\LARGE \textbf{Quasi-dynamic Traffic Light Control for a Single
Intersection}}}
\author{\textbf{Yanfeng Geng }and \textbf{Christos G. Cassandras}%
\thanks{{\footnotesize The authors' work is supported in part by NSF under
Grants EFRI-0735974 and CNS-1239021, by AFOSR under grant FA9550-12-1-0113, by
ONR under grant N00014-09-1-1051, and by ARO under Grant W911NF-11-1-0227.}}\\Division of Systems Engineering\\and Center for Information and Systems Engineering\\Boston University, gengyf@bu.edu, cgc@bu.edu}
\date{}
\maketitle

\begin{abstract}
We address the traffic light control problem for a single intersection by
viewing it as a stochastic hybrid system and developing a Stochastic Flow
Model (SFM) for it. We adopt a quasi-dynamic control policy based on partial
state information defined by detecting whether vehicle backlog is above or
below a certain threshold, without the need to observe an exact vehicle count.
The policy is parameterized by green and red cycle lengths which depend on
this partial state information. Using Infinitesimal Perturbation Analysis
(IPA), we derive on-line gradient estimators of an average traffic congestion
metric with respect to these controllable green and red cycle lengths when the
vehicle backlog is above or below the threshold. The estimators are used to
iteratively adjust light cycle lengths so as to improve performance and, in
conjunction with a standard gradient-based algorithm, to seek optimal values
which adapt to changing traffic conditions. Simulation results are included to
illustrate the approach and quantify the benefits of quasi-dynamic traffic
light control over earlier static approaches.

\end{abstract}

\section{Introduction}

The Traffic Light Control (TLC) problem aims at dynamically controlling the
flow of traffic at an intersection through the timing of green/red light
cycles with the objective of reducing congestion, hence also the delays
incurred by drivers. The more general problem involves a set of intersections
and traffic lights with the objective of reducing overall congestion over an
area covering multiple urban blocks. Control strategies employed for TLC
problems are generally classified into two categories: \emph{fixed-cycle
strategies} and \emph{traffic-responsive strategies}. Fixed-cycle Strategies
are derived off-line based on historical constant demands and turning rates
for each stream; traffic-responsive strategies make use of real-time
measurements to calculate in real time the best signal settings
\cite{Papageorgiou03}. Recent technological developments involving better,
inexpensive sensors and wireless sensor networks have enabled the collection
of data (e.g., counting vehicles in a specific road section) which can be used
for traffic-responsive strategies. Thus, methodologies that would not be
possible to implement not long ago are now becoming feasible. The approach
proposed in this paper to the TLC problem is specifically intended to exploit
these recent developments.

Numerous algorithms have been proposed to solve the TLC problem. It is
formulated as a Mixed Integer Linear Programming (MILP) problem in
\cite{Dujardin11}, and as an Extended Linear Complementary Problem (ELCP) in
\cite{Schutter99}. A Markov Decision Process (MDP) approach has been proposed
in \cite{Yu06} and Reinforcement Learning (RL) was used in \cite{Thorpe97}. A
game theoretic viewpoint is given in \cite{Alvarez10}. Due to its complexity
when viewed as an optimization problem, fuzzy logic is often used in both a
single (isolated) junction \cite{Murat05} and multiple junctions
\cite{Choi02}. The authors in \cite{Porche99} proposed an ALLONS-D framework
(Adaptive Limited Lookahead Optimization of Network Signals - Decentralized),
which is a decentralized method based on the Rolling Horizon (RH) concept.
Perturbation analysis techniques were used in \cite{Head96} and a formal
approach using Infinitesimal Perturbation Analysis (IPA) to solve the TLC
problem was presented in \cite{Panayiotou05} for a single intersection.

In \cite{GengCDC12}, we studied the TLC problem for a single intersection
using a Stochastic Flow Model (SFM) and Infinitesimal Perturbation Analysis
(IPA), and extended the method to multiple intersections in \cite{GengWODES12}%
. In this prior work, the green/red cycle lengths are viewed
as controllable parameters. The traffic light controller adjusts their values
based on data collected over an interval at the end of which an IPA estimator
dictates the adjustments. The data consist of counters and timers for simple,
directly observable events, but no state information (in the form of
instantaneous vehicle backlogs) is used. In this paper, we make a first
attempt to use state feedback for the controller in between two light cycle
adjustment points and use this information to improve the adjustments made.
However, since it is unrealistic to obtain instantaneous vehicle backlog
information and derive a fully \emph{dynamic} controller, we make use of
partial state information and derive a \emph{quasi-dynamic} controller. In
particular, we define for each traffic flow $i$ a minimum and maximum green
light cycle length, $\theta_{i,1}$, $\theta_{i,2}$ respectively, and allow
light switches as long as the cycle has exceeded $\theta_{i,1}$ depending
\emph{only} on whether some threshold of the vehicle backlog is reached,
assuming that such events are observable. We use IPA to estimate the
sensitivities of an average traffic congestion metric with respect to these
parameters, hence improving and seeking to optimize overall performance.

In our analysis, we still adopt a stochastic hybrid system modeling framework
\cite{Cassandras08},\cite{Cassandras06}, since the problem involves both
event-driven dynamics in the switching of traffic lights and time-driven
dynamics that capture the flow of vehicles through an intersection. A SFM as
introduced in \cite{Cassandras02} treats flow models as stochastic processes.
In the TLC problem, this is consistent with continuously and randomly varying
traffic flows, especially in heavy traffic conditions. With only minor
technical assumptions imposed on the properties of such processes, a general
IPA theory for stochastic hybrid systems was recently presented in
\cite{Wardi10},\cite{Cassandras10} through which one can estimate on line
gradients of certain performance measures with respect to various controllable
parameters. These estimates may be incorporated in standard gradient-based
algorithms to optimize system parameter settings. IPA estimates become biased
when dealing with aspects of queueing systems such as multiple user classes,
blocking due to limited resource capacities, and various forms of feedback
control. The use of IPA in stochastic hybrid systems, however, circumvents
these limitations and yields simple unbiased gradient estimates (under mild
technical conditions) of useful metrics (see \cite{Cassandras10}.) We
emphasize that the IPA gradient estimates we derive in the TLC are independent
of the stochastic characteristics of all the vehicle traffic flows involved,
rendering them robust to traffic variations and requiring no explicit models
for the traffic flows.

The rest of this paper is organized as follows. In Section \ref{sec2}, we
formulate the TLC problem for two intersections and construct a SFM. In
Section III, we derive an IPA estimator for a cost function gradient with
respect to a controllable parameter vector defined by green and red cycle
lengths. Simulation-based examples are given in Section IV and we conclude
with Section V.

\section{Problem Formulation}

\label{sec2} 
\begin{figure}[tbh]
\centering
\includegraphics[scale = 0.4]{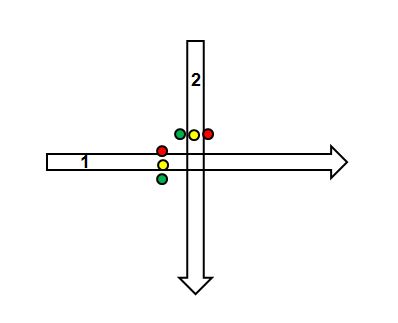} \caption{A single traffic
intersection}%
\label{junction}%
\end{figure}

A single isolated intersection is shown in Fig. \ref{junction},
where there are two roads and two traffic lights, with each traffic light
controlling the\ associated incoming traffic flow. For simplicity, we make the
following assumptions: $(i)$ Left-turn and right-turn traffic flows are not
considered, i.e., traffic lights only control vehicles going straight. $(ii)$
A YELLOW light is combined with a RED light (therefore, the YELLOW light
duration is not explicitly controlled).

\subsection{Quasi-dynamic Control}

We assign to each queue $i$ a guaranteed minimum GREEN cycle length
$\theta_{i,1}$, and a maximum length $\theta_{i,2}$. These are both
controllable parameters so that the controllable parameter vector of interest
is $\theta=[\theta_{1,1},\theta_{1,2},\theta_{2,1},\theta_{2,2}]$. We define a
state vector $x(\theta,t)=[x_{1}(\theta,t),x_{2}(\theta,t)]$ where
$x_{i}(\theta,t)\in\mathbb{R}^{+}$ is the content of queue $i$. For each queue
$i$, we also define a left-continuous \textquotedblleft
clock\textquotedblright\ state variable $z_{i}(\theta,t)$, $i=1,2$, which
measures the time since the last switch from RED to GREEN of the traffic light
for queue $i$, therefore,  $z_{i}(\theta,t)\in\lbrack0,\theta_{i,2}]$. Setting
$z(\theta,t)=[z_{1}(\theta,t),z_{2}(\theta,t)]$, the complete system state
vector is $[x(\theta,t),z(\theta,t)]$. For notational simplicity, we will
write $x_{i}(t)$, $z_{i}(t)$ when no confusion arises. 

At any time $t$, the feasible control set for the traffic
light controller is $U=\{1,2\}$, where we define the control as
\begin{equation}
u(x(t),z(t))\equiv\left\{
\begin{array}
[c]{ll}%
1 & \text{ set road 1 GREEN, road 2 RED}\\
2 & \text{ set road 2 GREEN, road 1 RED}%
\end{array}
\right.  \label{uxz}%
\end{equation}
A \emph{dynamic} controller is one that makes full use of the state
information $z(t)$ and $x(t)$. Obviously, $z(t)$ is the controller's known
internal state, but the queue content state is generally not observable. We
assume, however, that it is partially observable. Specifically, we can only
observe whether $x_{i}(t)$ is below or above some given threshold
$S_{i},i=1,2$ (this is consistent with actual traffic systems where a single
sensor (typically, an inductive loop detector) is installed at each road near
the intersection). Based on such partially observed states, we partition the
queue content state space into four regions as follows:
\[
X_{0}=\{(x_{1},x_{2}):x_{1}(t)<S_{1},\text{ }x_{2}(t)<S_{2}\}
\]%
\[
X_{1}=\{(x_{1},x_{2}):x_{1}(t)<S_{1},\text{ }x_{2}(t)\geq S_{2}\}
\]%
\[
X_{2}=\{(x_{1},x_{2}):x_{1}(t)\geq S_{1},\text{ }x_{2}(t)<S_{2}\}
\]%
\[
X_{3}=\{(x_{1},x_{2}):x_{1}(t)\geq S_{1},\text{ }x_{2}(t)\geq S_{2}\}
\]
The \emph{quasi-dynamic}\ controller we consider is of the form $u(z(t),X(t))$%
, where $X(t)=X_{0},\ldots,X_{3}$, and is defined as follows. If
$X(t)\in\{X_{0},X_{3}\}$,
\begin{equation}
u(z(t))=\left\{
\begin{array}
[c]{ll}%
1 & \text{if }z_{1}(t)\in(0,\theta_{1,2}),z_{2}(t)=0\\
2 & \text{otherwise}%
\end{array}
\right.  \label{u0}%
\end{equation}
If $X(t)=X_{1}$,
\begin{equation}
u(z(t))=\left\{
\begin{array}
[c]{ll}%
1 & \text{if }z_{1}(t)\in(0,\theta_{1,1}),z_{2}(t)=0\\
2 & \text{otherwise}%
\end{array}
\right.  \label{u1}%
\end{equation}
If $X(t)=X_{2}$,
\begin{equation}
u(z(t))=\left\{
\begin{array}
[c]{ll}%
2 & \text{if }z_{2}(t)\in(0,\theta_{2,1}),z_{1}(t)=0\\
1 & \text{otherwise}%
\end{array}
\right.  \label{u2}%
\end{equation}
This is a simple form of hysteresis control satisfying the following simple
rules with $j\neq i$:

\begin{itemize}
\item If the GREEN light cycle at queue $i$ reaches $\theta_{i,2}$, it
switches to RED.

\item If the GREEN light cycle at queue $i$ reaches $\theta_{i,1}$, and
$x_{i}<S_{i}$ (low $i$ traffic), $x_{j}\geq S_{j}$ (high $j$ traffic), it
switches to RED.

\item If the GREEN light cycle at queue $i$ has exceeded $\theta_{i,1}$, and
$x_{i}$ decreases below $S_{i}$ but $x_{j}\geq S_{j}$, it switches to RED.

\item If the GREEN light cycle at queue $i$ has exceeded $\theta_{i,1}$, and
$x_{j}$ increases above $S_{j}$ but $x_{i}<S_{i}$, it switches to RED.
\end{itemize}

Clearly, the GREEN light cycle may now be dynamically interrupted anytime
after $\theta_{i,1}$ based on the partial state feedback provided through
$X(t)$. Let $\tau_{k}$ be the $k$th time instant when a GREEN cycle starts at
queue $i$ and let $j\neq i$. Define
\[
r_{i}(\tau_{k})=\min\{t\arrowvert t\geq\tau_{k},z_{i}(t)\geq\theta_{i1}%
,x_{i}(t)<S_{i},x_{j}(t)\geq S_{j}\}
\]
This is the earliest time when all conditions are satisfied to interrupt a
GREEN cycle: the minimum GREEN cycle length $\theta_{i1}$ has been reached,
the queue $i$ length is low, and the competing queue $j$ length is high. At
$t=r_{i}(\tau_{k})$, a GREEN to RED light switching event is forced as long as
the residual GREEN cycle satisfies $\theta_{i,2}-r_{i}(\tau_{k})> 0$.
Therefore, the condition for such an interruption event is%
\[
r_{i}(\tau_{k})<\theta_{i,2}%
\]
At $t=\theta_{i,2}$, the GREEN light is forced to switch to RED.
This is an essential component of the quasi-dynamic controller we have defined. It will also be reflected by the state dynamics in the next subsection.

\subsection{System Dynamics}

The system, as described above, involves a number of stochastic processes
which are all defined on a common probability space $(\Omega,F,P)$. Each of
the two roads is considered as a queue with a random \emph{arrival} flow
process $\{\alpha_{n}(t)\},n=1,2.$, where $\alpha_{n}(t)$ is the instantaneous
vehicle arrival rate at time $t$. When the traffic light corresponding to road
$n$ is GREEN, the \emph{departure} flow process is denoted by $\{\beta
_{n}(t)\},n=1,2$. We emphasize again that the IPA estimators we will derive do
not require any knowledge of the processes $\{\alpha_{n}(t)\}$ and
$\{\beta_{n}(t)\}$; only estimates of arrival and departure flows at specific
observable event times are involved. As in prior work \cite{Cassandras10}%
,\cite{GengCDC12}, we assume only that $\alpha_{n}(t)$ and $\beta_{n}(t)$ are
piecewise continuous w.p. 1.

We can now write the dynamics of each state variable $z_{i}(t)$ based on the
control policy (\ref{uxz})-(\ref{u2}) as follows:%
\begin{align}
\label{switchRule}\dot{z}_{i}(t) &  =\left\{
\begin{array}
[c]{ll}%
1 & \text{if }z_{j}(t)=0,\text{ }j\neq i\\
0 & \text{otherwise}%
\end{array}
\right.  \\
z_{i}(t^{+}) &  =0\text{ if }z_{i}(t)=\theta_{i,2}\\
&  \text{ or }z_{i}(t)=\theta_{i,1},\text{ }x_{i}(t)<S_{i},\text{ }%
x_{j}(t)\geq S_{j}\nonumber\\
&  \text{ or }z_{i}(t)>\theta_{i,1},x_{i}(t^{-})>S_{i},x_{i}(t^{+}%
)=S_{i},x_{j}(t)\geq S_{j}\nonumber\\
&  \text{ or }z_{i}(t)>\theta_{i,1},x_{i}(t)<S_{i},x_{j}(t^{-})<S_{j}%
,x_{j}(t^{+})=S_{j}\nonumber
\end{align}
Note that $z_{i}(t)$ is reset to $0$ as soon as a GREEN light switches to RED
and it remains at this value while the light is GREEN for queue $j\neq i$.

The dynamics of each state variable $x_{n}(t)$ are as follows:%
\begin{equation}
\dot{x}_{n}(t)=\left\{
\begin{array}
[c]{ll}%
\alpha_{n}(t) & \text{if }z_{n}(t)=0\\
0 & \text{if }x_{n}(t)=0\text{ and }\alpha_{n}(t)\leq\beta_{n}(t)\\
\alpha_{n}(t)-\beta_{n}(t) & \text{otherwise}%
\end{array}
\right.  \label{dxdt}%
\end{equation}

\begin{figure}[h]
\centering
\includegraphics[scale = 0.37]{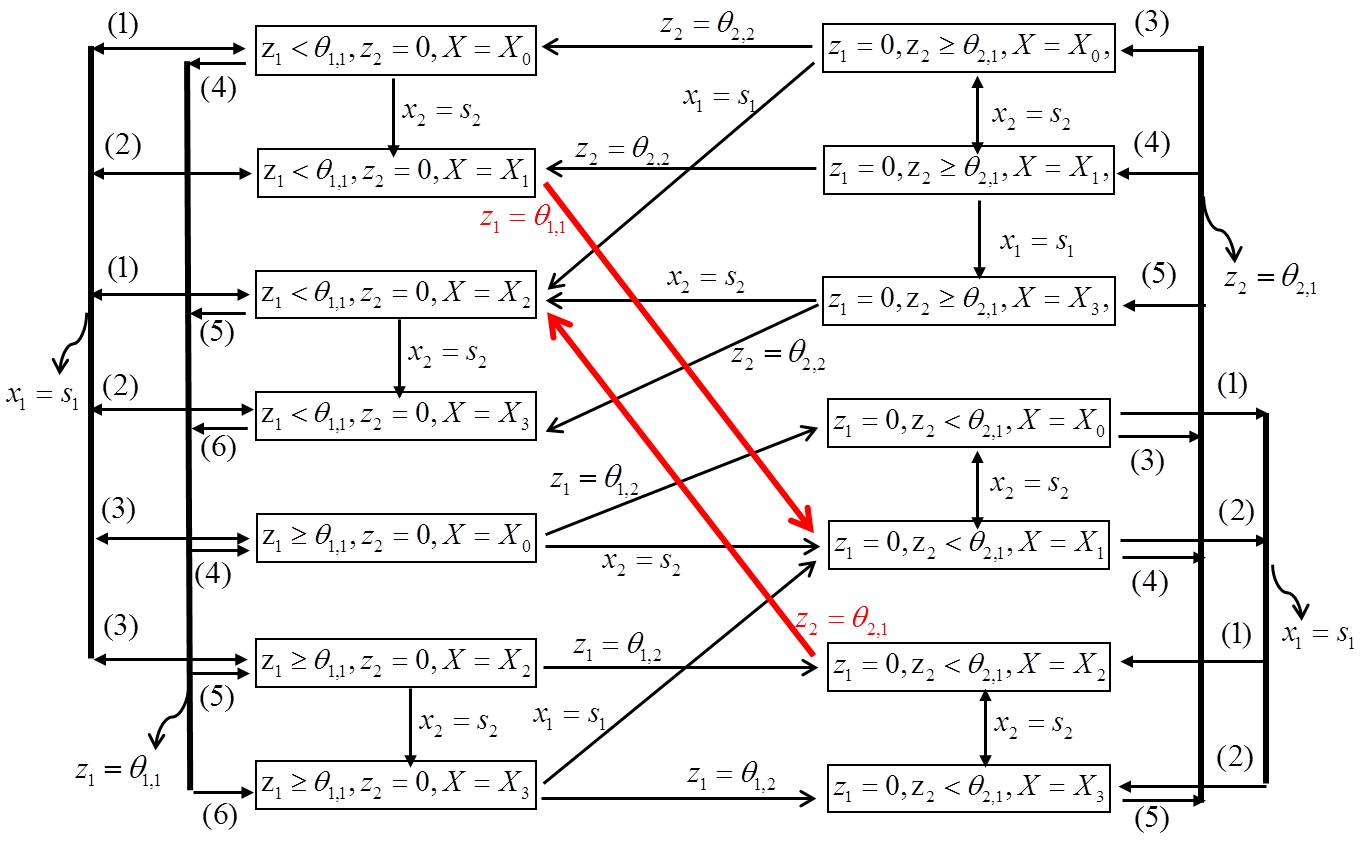} \caption{Stochastic Hybrid
Automaton under Quasi-dynamic Control}%
\label{Automaton}%
\end{figure}

Using the standard definition of a Stochastic Hybrid Automaton (SHA) (e.g.,
see \cite{Cassandras08}), we have a SHA for the system as shown in Fig.
\ref{Automaton}. To simplify the automaton, we omit the dynamics of $x_{n}(t)$
and $z_{n}(t)$ and aggregate the states $x_{n}(t)=0$ and $x_{n}(t)>0$ as one
state. As we can see, the system has 14 modes, which are defined by different
combinations of $x_{n}(t)$ and $z_{n}(t)$. Two transient modes are not shown
in the SHA: $z_{1}\geq\theta_{1,1},z_{2}=0,x_{1}<S_{1},x_{2}\geq S_{2}$ and
$z_{1}=0,z_{2}\geq\theta_{2,1},x_{1}\geq S_{1},x_{2}<S_{2}$. It is easy to show that the control policies (\ref{u1})-(\ref{u2}) force
irreversible transitions into recurrent states in Fig. \ref{Automaton}.

Next, we define the set of all events in this system (causing all mode
transitions) as $\Phi_{n}=\{e_{1},e_{2},e_{3},e_{4},e_{5},e_{6},e_{7}\}$ where
$e_{1}$ is the guard condition $[x_{n}=S_{n}$ from below$]$; $e_{2}$ is the
guard condition $[x_{n}=S_{n}$ from above$]$; $e_{3}$ is the guard condition
$[z_{i}=\theta_{i,1}]$, i.e., the GREEN cycle length reaches its lower limit;
$e_{4}$ is the guard condition $[z_{i}=\theta_{i,2}]$, i.e., the GREEN cycle
length reaches its upper limit; $e_{5}$ is the guard condition $[x_{n}=0$ from
above$]$, i.e., the $n$th queue becomes empty; $e_{6}$ is a switch in the sign
of $\alpha_{n}(t)-\beta_{n}(t)$ from non-positive to strictly positive; and
$e_{7}$ is a switch in the sign of $\alpha_{n}(t)$ from $0$ to strictly
positive. Events $e_{5}$, $e_{6}$ and $e_{7}$ are not shown in the above
automaton (see \cite{GengCDC12} for detailed descriptions). 

\begin{figure}[tbh]
\centering
\includegraphics[scale = 0.40]{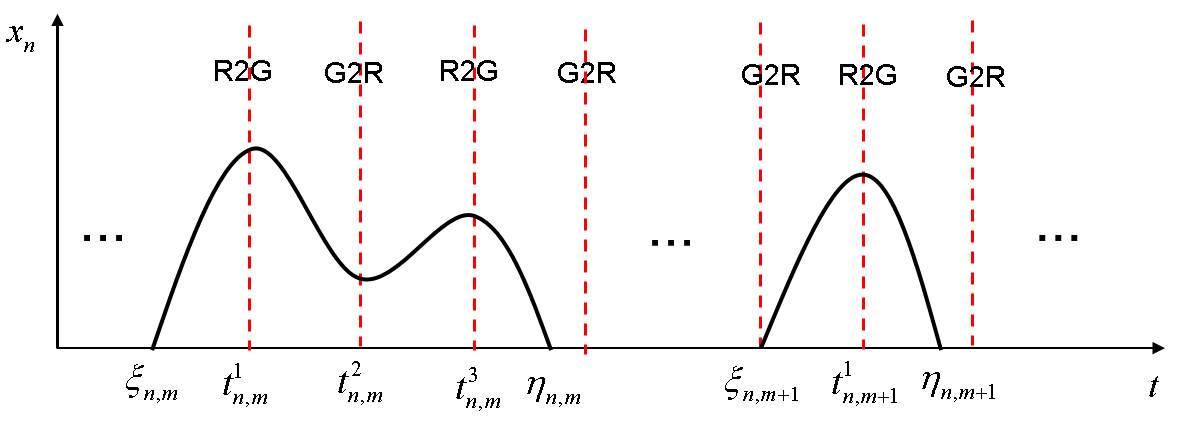}\caption{A typical sample path
of a traffic light queue}%
\label{SamplePath}%
\end{figure}

A typical sample path of any one of the queue content states (as shown in Fig.
\ref{SamplePath}) consists of intervals over which $x_{n}(t)>0$, which we call
\emph{Non-Empty Periods} (NEPs), followed by intervals where $x_{n}(t)=0$,
which we call \emph{Empty Periods} (EPs). Thus, the entire sample path
consists of a series of alternating NEPs and EPs. For easier reference, we let
\textquotedblleft E\textquotedblright\ denote any \textquotedblleft NEP
end\textquotedblright\ event (caused by $e_{5}$), \textquotedblleft R2G\textquotedblright\ denote a
light switching event from RED to GREEN, \textquotedblleft
G2R\textquotedblright\ denote a light switching event from GREEN to RED (both
G2R and R2G are caused by $e_{1},\ldots,e_{4}$), and \textquotedblleft
S\textquotedblright\ denote any \textquotedblleft NEP start\textquotedblright%
\ event, which is caused by $e_{6}$, $e_{7}$ or G2R (see also \cite{GengCDC12}%
). Our analysis will be based on studying these four event types.

In Fig. \ref{SamplePath}, the $m$th NEP in a sample path of any queue,
$m=1,2,\ldots$, is denoted by $[\xi_{n,m},\eta_{n,m})$, i.e., $\xi_{n,m}$,
$\eta_{n,m}$ are the occurrence times of the $m$th S\ and E\ event
respectively at this queue. During the $m$th NEP, $t_{n,m}^{j}$,
$j=1,\ldots,J_{m}$, denotes the time when a traffic light switching event
occurs (either R2G or G2R).

\subsection{Objective Function}

Our objective is to select $\theta$ so as to minimize a cost function that
measures a weighted mean of the queue lengths over a fixed time interval
$[0,T]$. Note that the threshold parameters $S_{i},i=1,2$, are assumed to be
fixed for the purpose of this paper. In particular, we define the sample
function%
\begin{equation}
L(\theta;x(0),z(0),T)=\frac{1}{T}\sum\limits_{n=1}^{2}\int\nolimits_{0}%
^{T}w_{n}x_{n}(\theta,t)dt\label{costfun1}%
\end{equation}
where $w_{n}$ is a cost weight associated with queue $n$ and $x(0),z(0)$ are
given initial conditions. It is obvious that since $x_{n}(t)=0$ during EPs of
queue $n$, we can rewrite (\ref{costfun1}) in the form
\begin{equation}
L(\theta;x(0),z(0),T)=\frac{1}{T}\sum\limits_{n=1}^{2}\sum\limits_{m=1}%
^{M_{n}}\int\nolimits_{\xi_{n,m}}^{\eta_{n,m}}w_{n}x_{n}(\theta
,t)dt\label{costfun}%
\end{equation}
where $M_{n}$ is the total number of NEPs during the sample path of queue $n$.
For convenience, we also define%
\begin{equation}
L_{n,m}(\theta)=\int\nolimits_{\xi_{n,m}}^{\eta_{n,m}}x_{n}(\theta
,t)dt\label{Lnm}%
\end{equation}
to be the sample cost associated with the $m$th NEP of queue $n$. We can now
define our overall performance metric as%
\begin{equation}
J(\theta;x(0),z(0),T)=E\left[  L(\theta;x(0),z(0),T\right]  \label{costfun3}%
\end{equation}
Recall that we do not impose any limitations on the processes $\{\alpha
_{n}(t)\}$ and $\{\beta_{n}(t)\}$ and only assume that $\alpha_{n}(t)$,
$\beta_{n}(t)$ are piecewise continuous w.p. 1. Thus, it is infeasible to
obtain a closed-form expression of $J(\theta;x(0),z(0),T)$. The value of IPA,
as developed for general stochastic hybrid systems in \cite{Cassandras10}, is
in providing the means to estimate the performance metric gradient $\nabla
J(\theta)$, by evaluating the sample gradient $\nabla L(\theta)$. As shown
elsewhere (e.g., see \cite{Cassandras10}), these estimates are unbiased under
mild technical conditions. Moreover, an important property of IPA estimates is
that they are often independent of the unknown processes $\{\alpha_{n}(t)\}$
and $\{\beta_{n}(t)\}$ or they depend on values of $\alpha_{n}(t)$ or
$\beta_{n}(t)$ at specific event times only. Such robustness properties of IPA
(formally established in \cite{YaoCgc11}) make it attractive for estimating on
line performance sensitivities with respect to controllable parameters such as
$\theta$ in our case. One can then use this information to either improve
performance or, under appropriate conditions, solve an optimization problem
and determine an optimal $\theta^{\ast}$ through an iterative scheme:
\begin{equation}
\theta_{i,k+1}=\theta_{i,k}-\gamma_{k}H_{i,k}(\theta_{k},x(0),T,\omega
_{k}),k=0,1,...\label{iteration}%
\end{equation}
where $H_{i,k}(\theta_{k},x(0),T,\omega_{k})$ is an estimate of $dJ/d\theta
_{i}$ based on the information obtained from the sample path denoted by
$\omega_{k}$, and $\gamma_{k}$ is the stepsize at the $k$th iteration. Next we
will focus on how to obtain $dL/d\theta$. We may then also obtain
$\theta^{\ast}$ through (\ref{iteration}), provided that $\{\alpha_{n}(t)\}$
and $\{\beta_{n}(t)\}$ are stationary.

\section{Infinitesimal Perturbation Analysis (IPA)}

To simplify notation,, we redefine $\theta_{1,1}, \theta_{1,2}$ as $\theta
_{1}$, $\theta_{2}$ and $\theta_{2,1}, \theta_{2,2}$ as $\theta_{3}$,
$\theta_{4}$, and define the derivatives of the states $x_{n}(t,\theta)$ and
$z_{i}(t,\theta)$ and event times $\tau_{k}(\theta)$ with respect to
$\theta_{i}$, $i=1,...,4$, as follows:
\begin{equation}
x_{n,i}^{\prime}(t)\equiv\frac{\partial x_{n}(\theta,t)}{\partial\theta_{i}%
},\text{ }z_{i,i}^{\prime}(t,\theta)\equiv\frac{\partial z_{i}(\theta
,t)}{\partial\theta_{i}},\text{ }\tau_{k,i}^{\prime}\equiv\frac{\partial
\tau_{k}(\theta)}{\partial\theta_{i}} \label{IPAnotation}%
\end{equation}

Consider a sample path of the system as modeled in Fig. \ref{Automaton} over
$[0,T]$ and let $\tau_{k}(\theta)$ denote the occurrence time of the $k$th
event (of any type), where we stress its dependence on $\theta$. Taking
derivatives with respect to $\theta_{i}$ in (\ref{costfun}), and observing
that $x_{n}(\xi_{n,m})=x_{n}(\eta_{n,m})=0$, we obtain%
\begin{align}
\frac{dL(\theta)}{d\theta_{i}} = \frac{1}{T}\sum\limits_{n=1}^{2}%
\sum\limits_{m=1}^{M_{n}}w_{n}\frac{dL_{n,m}(\theta)}{d\theta_{i}}\label{dLi}%
\end{align}

Observe that the determination of the sample derivatives depends on the state
derivatives $x_{n,i}^{\prime}(t)$. The purpose of IPA is to evaluate these
derivatives as functions of observable sample path quantities. We pursue this
next, using the framework established in \cite{Cassandras10} where, for
arbitrary stochastic hybrid systems, it is shown that the state and event time
derivatives in (\ref{IPAnotation}) can be obtained from three fundamental
\textquotedblleft IPA equations\textquotedblright. For the sake of
self-sufficiency, these equations are rederived here as they pertain to our
specific SFM.

\subsection{IPA review}

Looking at (\ref{dxdt}), note that the dynamics of $x_{n}(t)$ are fixed over
any interevent interval $[\tau_{k},\tau_{k+1})$ and we write $\dot{x}%
_{n}(t)=f_{n,k}(t)$ to represent the appropriate expression on the
right-hand-side of (\ref{dxdt}) over this interval. We have $x_{n}%
(t)=x_{n}(\tau_{k})+\int\nolimits_{\tau_{k}}^{t}f_{n,k}(\tau)d\tau$. Taking
derivatives with respect to $\theta_{i}$ and letting $t=\tau_{k}^{+}$, we
obtain
\begin{equation}
x_{n,i}^{\prime}(\tau_{k}^{+})=x_{n,i}^{\prime}(\tau_{k}^{-})+[f_{n,k-1}%
(\tau_{k}^{-})-f_{n,k}(\tau_{k}^{+})]\cdot\tau_{k,i}^{\prime}\label{jumps}%
\end{equation}
Moreover, further taking derivatives with respect to $t$, we get, for all
$t\in\lbrack\tau_{k},\tau_{k+1})$,
\begin{equation}
\frac{d}{dt}x_{n,i}^{\prime}(t)=\frac{\partial f_{n,k}}{\partial x_{n}%
}(t)x_{n,i}^{\prime}(t)+\frac{\partial f_{n,k}}{\partial\theta_{i}%
}(t)\label{dxdtdtheta}%
\end{equation}
Since $\frac{\partial f_{n,k}}{\partial x_{n}}=\frac{\partial f_{n,k}%
}{\partial\theta_{i}}=0$ and we get $\frac{d}{dt}x_{n,i}^{\prime}(t)=0$.
Therefore, $x_{n,i}^{^{\prime}}(t)$ remains constant over all $t\in\lbrack
\tau_{k},\tau_{k+1})$:
\begin{equation}
x_{n,i}^{\prime}(t)=x_{n,i}^{\prime}(\tau_{k}^{+}),\text{ \ \ \ }t\in
\lbrack\tau_{k},\tau_{k+1})\label{xprime(t)}%
\end{equation}
Thus, focusing on a NEP of $x_{n}(t)$, the queue content derivative is
piecewise constant with jumps occurring according to (\ref{jumps}). The next step
is to obtain the event time derivatives $\tau_{k,i}^{\prime}$ appearing in
(\ref{jumps}).

Clearly $\tau_{k,i}^{\prime}$ depends on the type of event occurring at time
$\tau_{k}$. Following the framework in \cite{Cassandras10}, there are three
types of events for a general stochastic hybrid system. $(i)$ \emph{Exogenous
Events.} These events cause a discrete state transition independent of
$\theta$ and satisfy $\tau_{k,i}^{\prime}=0$. $(ii)$ \emph{Endogenous
Events.} Such an event occurs at time $\tau_{k}$ if there exists a
continuously differentiable function $g_{k}:\mathbb{R}^{n}\times
\Theta\rightarrow\mathbb{R}$ such that $\tau_{k}\ =\ \min\{t>\tau
_{k-1}\ :\ g_{k}\left(  x\left(  \theta,t\right)  ,\theta\right)  =0\}$, where
the function $g_{k}$ normally corresponds to a guard condition in a hybrid
automaton. Taking derivatives with respect to $\theta_{i}$, $i=1,\ldots,m$, it
is straightforward to obtain
\begin{equation}
\tau_{k,i}^{\prime}=-\frac{\frac{\partial g_{k}}{\partial\theta_{i}}%
+\sum_{j=1}^{N}\frac{\partial g_{k}}{\partial x_{j}}x_{j,i}^{\prime}(\tau
_{k}^{-})}{\sum_{j=1}^{N}\frac{\partial g_{k}}{\partial x_{j}}f_{j,k-1}%
(\tau_{k}^{-})} \label{dtaudtheta}%
\end{equation}
$(iii)$ \emph{Induced Events.} Such an event occurs at time $\tau_{k}$ if it
is triggered by the occurrence of another event at time $\tau_{m}\leq\tau_{k}%
$. (details can be found in \cite{Cassandras10}.)

\subsection{State and event time derivatives}

In the following, we consider each of the four event types (E, S, R2G, G2R)
for queue $n$ that were identified in the previous section and derive the
corresponding event time derivatives. Based on these, we can then also derive
the state derivatives through (\ref{jumps}) and (\ref{xprime(t)}).

\textbf{(1)} \emph{Event E ending a NEP}. This is an endogenous event that
occurs when $x_{n}(\theta,t)=0$. Thus, when such an event occurs at $\tau_{k}%
$, let $g_{k}(x(\theta,t),\theta)=x_{n}(\theta,t)=0$. Using (\ref{dtaudtheta}%
), we get $\tau_{k,i}^{\prime}=\frac{-x_{n,i}^{\prime}(\tau_{k}^{-}%
)}{f_{n,k-1}(\tau_{k}^{-})}$. Looking at (\ref{dxdt}), note that
$f_{n,k-1}(\tau_{k}^{-})=\alpha_{n}(\tau_{k})-\beta_{n}(\tau_{k})$ and
$f_{n,k}(\tau_{k}^{+})=0$. Using these values in (\ref{jumps}) along with
$\tau_{k,i}^{\prime}$ above we get%
\[
x_{n,i}^{\prime}(\tau_{k}^{+})=x_{n,i}^{\prime}(\tau_{k}^{-})-\frac
{[\alpha_{n}(\tau_{k})-\beta_{n}(\tau_{k})]x_{n,i}^{\prime}(\tau_{k}^{-}%
)}{\alpha_{n}(\tau_{k})-\beta_{n}(\tau_{k})}=0
\]
Thus, at the end of a NEP $[\xi_{n,m},\eta_{n,m})$ of queue $n$ we have%
\begin{equation}
x_{n,i}^{\prime}(\eta_{n,m}^{+})=0 \label{E_event}%
\end{equation}
indicating that these state derivatives are always reset to $0$ upon ending a NEP.

\textbf{(2)} \emph{G2R event}. If a G2R event occurs within a NEP (i.e.,
$x_{n}(\tau_{k})>0$), then, based on (\ref{dxdt}), we have $f_{n,k-1}(\tau
_{k}^{-})=\alpha_{n}(\tau_{k})-\beta_{n}(\tau_{k})$ and $f_{n,k}(\tau_{k}%
^{+})=\alpha_{n}(\tau_{k})$. Therefore, from (\ref{jumps}) we get
$x_{n,i}^{\prime}(\tau_{k}^{+})=x_{n,i}^{\prime}(\tau_{k}^{-})-\beta_{n}%
(\tau_{k})\cdot\tau_{k,i}^{\prime}$. If, on the other hand, $x_{n}(\tau
_{k})=0$, then $f_{n,k-1}(\tau_{k}^{-})=0$ and we get $x_{n,i}^{\prime}%
(\tau_{k}^{+})=x_{n,i}^{\prime}(\tau_{k}^{-})-\alpha_{n}(\tau_{k})\cdot
\tau_{k,i}^{\prime}$. Combining these two results,
\begin{equation}
x_{n,i}^{\prime}(\tau_{k}^{+})=x_{n,i}^{\prime}(\tau_{k}^{-})-\left\{
\begin{array}
[c]{ll}%
\beta_{n}(\tau_{k})\cdot\tau_{k,i}^{\prime} & \text{if }x_{n}(\tau_{k})>0\\
\alpha_{n}(\tau_{k})\cdot\tau_{k,i}^{\prime} & \text{if }x_{n}(\tau_{k})=0
\end{array}
\right.  \label{type2di2}%
\end{equation}
Next we consider all four events $e_{1},\ldots,e_{4}$ which may trigger the
G2R event.

\textbf{Case (2a)}: G2R event at queue $i$ is triggered by $z_{i}=\theta
_{i,2}$ ($e_{4}$ event). This is an endogenous event. In the model studied in
\cite{GengCDC12}, G2R and R2G events alternate with fixed GREEN/RED cycles and
$\tau_{k,i}^{\prime}$ is obtained by simply counting traffic light switches.
However, under the quasi-dynamic control considered here, the GREEN cycle
length can take any value in $[\theta_{i,1},\theta_{i,2}]$ and the same method
no longer applies. In the following lemma we derive $\tau_{k,i}^{\prime}$.
\newtheorem{lemma}{Lemma} \begin{lemma} \label{lemma4}
Let $\tau_k$ be the time of a G2R event induced by $z_{n}=\theta_{n,2}, n=1,2$, and $\tau_p$ be the last R2G event time before $\tau_k$. We then have:
\begin{equation}
\tau_{k,i}^{\prime}=\left\{
\begin{array}
[c]{ll}%
1 + \tau_{p,i}^{\prime} & \text{if } i=2,n=1 \text{ or } i=4,n=2\\
\tau_{p,i}^{\prime} & otherwise \label{2a}
\end{array}
\right.
\end{equation}
\end{lemma}

\emph{Proof}: Without loss of generality, we use queue 1, i.e., $n=1$ above,
to derive $\tau_{k,i}^{\prime}=\frac{\partial\tau_{k}(\theta)}{\partial
\theta_{i}}$ for all $i=1,2,3,4$. Setting
\begin{equation}
g_{k}(x(\theta,t),\theta)=z_{1}-\theta_{2}=0\label{z1theta2}%
\end{equation}
in (\ref{dtaudtheta}) and taking derivatives with respect to $\theta_{1}$
above, we have
\begin{equation}
\tau_{k,1}^{\prime}=-\left(  \dot{z}_{1}(\tau_{k}^{-})\right)  ^{-1}\left(
z_{1,1}^{^{\prime}}(\tau_{k}^{-})-0\right)  =-z_{1,1}^{^{\prime}}(\tau_{k}%
^{-})\label{tauprime_k1}%
\end{equation}
where $\dot{z}_{1}(\tau_{k}^{-}))=1$ since
\[
\dot{z}_{1}(t)=f_{1,k-1}(t)=\left\{
\begin{array}
[c]{cc}%
1 & \text{if }z_{2}(t)=0\\
0 & \text{otherwise}%
\end{array}
\right.
\]
By the definition of $\tau_{p}$, for any $t\in\left[  \tau_{p},\tau
_{k}\right)  $, $f_{1}(t)=\dot{z}_{1}(t)=1$. Using (\ref{dxdtdtheta}), we get
\begin{equation}
\frac{d}{dt}z_{1,1}^{^{\prime}}(t)=0\label{ddtz}%
\end{equation}
so that $z_{1,1}^{^{\prime}}(t)$ remains constant for all $t\in\lbrack\tau
_{p},\tau_{k})$, i.e.,
\[
z_{1,1}^{^{\prime}}(\tau_{k}^{-})=z_{1,1}^{^{\prime}}(\tau_{p}^{+})
\]
At time $\tau_{p}$, according to (\ref{jumps}), we have
\begin{equation}
z_{1,1}^{^{\prime}}(\tau_{p}^{+})=z_{1,1}^{\prime}(\tau_{p}^{-})+[f_{1,p-1}%
(\tau_{p}^{-})-f_{1,p}(\tau_{p}^{+})]\cdot\tau_{p,1}^{^{\prime}}\label{z11p}%
\end{equation}
Define $\tau_{r}$ to be the last G2R event time before $\tau_{p}$. At
$\tau_{r}$, $z_{1}(\tau_{r}^{+})$ is reset to 0 according to (\ref{switchRule}%
), so that $z_{1,1}^{\prime}(\tau_{r}^{+})=0$. For any $t\in\left[  \tau
_{r},\tau_{p}\right)  $, we have $f_{1,p-1}(t)=\dot{z}_{1}(t)=0$, thus
$z_{1,1}^{\prime}(\tau_{p}^{-})=z_{1,1}^{\prime}(\tau_{r}^{+})=0$. The light
switches to GREEN after $\tau_{p}$, so we have $f_{1,p}(\tau_{p}^{+})=1$.
Hence, (\ref{z11p}) becomes:
\[
z_{1,1}^{^{\prime}}(\tau_{p}^{+})=-\tau_{p,1}^{^{\prime}}%
\]
Returning to (\ref{tauprime_k1}), we get
\begin{equation}
\tau_{k,1}^{^{\prime}}=-z_{1,1}^{^{\prime}}(\tau_{k}^{-})=-z_{1,1}^{^{\prime}%
}(\tau_{p}^{+})=\tau_{p,1}^{^{\prime}}\label{g2r_long_d1}%
\end{equation}
Following the same analysis, we have
\begin{equation}
\tau_{k,3}^{^{\prime}}=\tau_{p,3}^{^{\prime}}\label{g2r_long_d3}%
\end{equation}%
\begin{equation}
\tau_{k,4}^{^{\prime}}=\tau_{p,4}^{^{\prime}}\label{g2r_long_d4}%
\end{equation}
This leaves only $\tau_{k,2}^{^{\prime}}$ to consider. Taking derivatives with
respect to $\theta_{2}$ in (\ref{z1theta2}):
\[
\tau_{k,2}^{\prime}=-\left(  \dot{z}_{1}(\tau_{k}^{-})\right)  ^{-1}\left(
z_{1,2}^{\prime}(\tau_{k}^{-})-1\right)  =1-z_{1,2}^{^{\prime}}(\tau_{k}^{-})
\]
As in (\ref{ddtz}), for any $t\in\lbrack\tau_{p},\tau_{k}),$%
\[
\frac{d}{dt}z_{1,2}^{^{\prime}}(t)=0
\]
and we get $z_{1,2}^{^{\prime}}(\tau_{k}^{-})=z_{1,2}^{^{\prime}}(\tau_{p}%
^{+}).$ At time $\tau_{p}$, we have
\[
z_{1,2}^{^{\prime}}(\tau_{p}^{+})=z_{1,2}^{^{\prime}}(\tau_{p}^{-}%
)+[f_{1,p-1}(\tau_{p}^{-})-f_{1,p}(\tau_{p}^{+})]\cdot\tau_{p,2}^{^{\prime}%
}=-\tau_{p,2}^{^{\prime}}%
\]
and we finally get
\begin{equation}
\tau_{k,2}^{^{\prime}}=1-z_{1,2}^{^{\prime}}(\tau_{k}^{-})=1+\tau
_{p,2}^{^{\prime}}\label{g2r_long_d2}%
\end{equation}
$\blacksquare$

This result leaves the value of $\tau_{p,i}^{\prime}$ unspecified. In fact,
this is the time when an R2G event occurs, which is \textbf{Case (3)}
considered in the sequel where we shall derive explicit expressions for
$\tau_{p,i}^{\prime}$.

\textbf{Case (2b)}: G2R event at queue $i$ is triggered by $z_{i}=\theta
_{i,1}$, where $x_{i}<S_{i}$ and $x_{j}\geq S_{j}$ ($e_{3}$ event). We use a
similar lemma for this case.

\begin{lemma}\label{lemma5}
Let $\tau_k$ be the time of a G2R event induced by $z_{n}=\theta_{n,1},n=1,2$, and $\tau_p$ be the last R2G event before $\tau_k$. We  then have
\begin{equation}
\tau_{k,i}^{\prime}=\left\{
\begin{array}
[c]{ll}%
1 + \tau_{p,i}^{\prime} & \text{if } i=1, n=1 \text{ or }i=3,n=2 \\
\tau_{p,i}^{\prime} & otherwise\label{2b}
\end{array}
\right.
\end{equation}
\end{lemma}\emph{Proof}: The proof is similar to that of Lemma \ref{lemma4}
and is omitted.

\textbf{Case (2c)}: G2R event at queue $i$ is triggered by $x_{i}=S_{i}$ from
above, where $z_{i}>\theta_{i,1}$ and $x_{j}\geq S_{j}$ ($e_{2}$ event). This
is an endogenous event with $g_{k}(x(\theta,t),\theta)=x_{n}(\theta
,t)-S_{i}=0$. Using (\ref{dtaudtheta}), with $f_{n,k-1}(\tau_{k}^{-}%
)=\alpha_{n}(\tau_{k}^{-})-\beta_{n}(\tau_{k}^{-})$ we get
\begin{equation}
\tau_{k,i}^{\prime}=\frac{-x_{n,i}^{\prime}(\tau_{k}^{-})}{\alpha_{n}(\tau
_{k}^{-})-\beta_{n}(\tau_{k}^{-})}\label{g2r_i1}%
\end{equation}

\textbf{Case (2d)}: G2R event at queue $i$ is triggered by $x_{j}=S_{j}$ from
below, where $z_{i}>\theta_{i,1}$ and $x_{i}\leq S_{i}$ ($e_{1}$ event). This
is also an endogenous event with $g_{k}(x(\theta,t),\theta)=x_{j}%
(\theta,t)-S_{j}=0$. Using (\ref{dtaudtheta}), with ${f_{j,k-1}(\tau_{k}^{-}%
)}=\alpha_{j}(\tau_{k}^{-})$ we get
\begin{equation}
\tau_{k,i}^{\prime}=\frac{-x_{j,i}^{\prime}(\tau_{k}^{-})}{\alpha_{j}(\tau
_{k}^{-})}\label{g2r_i2}%
\end{equation}

\textbf{(3)} \emph{R2G event}. If the R2G event occurs within a NEP (i.e.,
$x_{n}(\tau_{p})>0$), then, based on (\ref{dxdt}), we have $f_{n,p-1}(\tau
_{p}^{-})=\alpha_{n}(\tau_{p})$ and $f_{n,p}(\tau_{p}^{+})=\alpha_{n}(\tau
_{p})-\beta_{n}(\tau_{p})$. From (\ref{jumps}) we get
\begin{equation}
x_{n,i}^{\prime}(\tau_{p}^{+})=x_{n,i}^{\prime}(\tau_{p}^{-})+\beta_{n}%
(\tau_{p})\cdot\tau_{p,i}^{\prime}\label{type3}%
\end{equation}
The derivation of $\tau_{p,i}^{\prime}$ is similar to \textbf{Case (2)} as
detailed next.

\textbf{Case (3a)}: R2G event at queue $i$ is triggered by $g_{p}%
(x(\theta,t),\theta)=z_{j}-\theta_{j,2}=0$ ($e_{3}$ event). This is an
endogenous event. We have a lemma similar to Lemma \ref{lemma4} whose proof is
therefore omitted:

\begin{lemma} \label{lemma6}
Let $\tau_p$ be the time of a R2G event induced by $z_{j}=\theta_{j,2},j=1,2$, and $\tau_r$ be the last G2R event before $\tau_p$. We then have
\begin{equation}
\tau_{p,i}^{\prime}=\left\{
\begin{array}
[c]{ll}%
1 + \tau_{r,i}^{\prime} & \text{if }i=2,j=1\text{ or }i=4,j=2 \\
\tau_{r,i}^{\prime} & otherwise\\\label{3a}
\end{array}
\right.
\end{equation}
\end{lemma}

\textbf{Case (3b)}: R2G event at queue $i$ is triggered by $z_{j}=\theta
_{j,1}$, where $x_{j}<S_{j}$ and $x_{i}\geq S_{i}$ ($e_{3}$ event). This is an
endogenous event and we have another lemma similar to Lemma \ref{lemma4} whose
proof is also omitted:

\begin{lemma}\label{lemma7}
Let $\tau_p$ be the time of a R2G event induced by $z_{j}=\theta_{j,1},j=1,2$, and $\tau_r$ be the last G2R event time before $\tau_p$. We then have
\begin{equation}
\tau_{p,i}^{\prime}=\left\{
\begin{array}
[c]{ll}%
1 + \tau_{r,i}^{\prime} & \text{if }i=1,j=1\text{ or }i=3,j=2 \\
\tau_{r,i}^{\prime} & otherwise\\\label{3b}
\end{array}
\right.
\end{equation}
\end{lemma}

\textbf{Case (3c)}: R2G event at queue $i$ is triggered by $x_{j}=S_{j}$ from
above, where $z_{j}>\theta_{j,1}$ and $x_{i}\geq S_{i}$ ($e_{2}$ event). This
is an endogenous event with $g_{p}(x(\theta,t),\theta)=x_{j}(\theta
,t)-S_{j}=0$. Using (\ref{dtaudtheta}) with  $f_{j,p-1}(\tau_{p}^{-}%
)=\alpha_{j}(\tau_{p}^{-})-\beta_{j}(\tau_{p}^{-})$, we get
\begin{equation}
\tau_{p,i}^{\prime}=\frac{-x_{j,i}^{\prime}(\tau_{p}^{-})}{\alpha_{j}(\tau
_{p}^{-})-\beta_{j}(\tau_{p}^{-})}\label{r2g_i1}%
\end{equation}

\textbf{Case (3d)}: R2G event at queue $i$ is triggered by $x_{i}=S_{i}$ from
below, where $z_{j}>\theta_{j,1}$ and $x_{j}\leq S_{j}$ ($e_{1}$ event). This
is an endogenous event with $g_{p}(x(\theta,t),\theta)=x_{n}(\theta
,t)-S_{i}=0$. Using (\ref{dtaudtheta}) with $f_{n,p-1}(\tau_{p}^{-}%
)=\alpha_{n}(\tau_{p}^{-})$, we get
\begin{equation}
\tau_{p,i}^{\prime}=\frac{-x_{n,i}^{\prime}(\tau_{p}^{-})}{\alpha_{n}(\tau
_{p}^{-})}\label{r2g_i2}%
\end{equation}

\textbf{(4)} \emph{Event S starting a NEP}. There are three possible cases to
consider as follows.

\textbf{Case (4a)}: \emph{A NEP starts right after a G2R event}. This is an
endogenous event and was analyzed in Case \textbf{(2)} with $x_{n}(\tau
_{k})=0$ in (\ref{type2di2}), i.e., $x_{n,i}^{\prime}(\tau_{k}^{+}%
)=x_{n,i}^{\prime}(\tau_{k}^{-})-\alpha_{n}(\tau_{k})\zeta_{n,k}$. Since
$x_{n,i}^{\prime}(\xi_{n,m}^{-})=x_{n,i}^{\prime}(\eta_{n,m-1}^{+}%
)=0\label{xprime=0}$, we get
\begin{equation}
x_{n,i}^{\prime}(\tau_{k}^{+})=-\alpha_{n}(\tau_{k})\tau_{k,i}^{\prime}
\label{4a}%
\end{equation}

\textbf{Case (4b)}:\emph{ A NEP starts while }$z_{i}=0$\emph{, }$z_{j}>0$.
This is an exogenous event occurring during a RED cycle for queue $n$ and is
due to a change in $\alpha_{n}(t)$ from a zero to a strictly positive value.
Therefore, $\tau_{k,i}^{\prime}=0$, and
\begin{equation}
x_{n,i}^{\prime}(\tau_{k}^{+})=0\label{type4bc}%
\end{equation}

\textbf{Case (4c)}: \emph{A NEP starts while }$z_{j}=0$\emph{, }$z_{i}>0$.
This is an exogenous event occurring during a GREEN cycle for queue $n$ due to
a change in $\alpha_{n}(t)$ or $\beta_{n}(t)$ that results in $\alpha
_{n}(t)-\beta_{n}(t)$ switching from a non-positive to a strictly positive
value. The analysis is the same as \textbf{Case (4b)}.

This completes the derivation of all state and event time derivatives required
to evaluate the sample performance derivative in (\ref{dLi}).

\subsection{Cost Derivative}

Using the definition of $L_{n,m}(\theta)$ in (\ref{Lnm}), note that we can
decompose (\ref{dLi}) for each $n$ into its NEPs and evaluate the derivatives
$dL_{n,m}(\theta)/d\theta_{i}$. By virtue of (\ref{xprime(t)}), $x_{n,i}%
^{\prime}(t)$ is piecewise constant during a NEP and its value changes only at
an event point $t_{n,m}^{j}$, $j=1,...,J_{n,m}$. Therefore, we have
\begin{align}
\frac{dL_{n,m}(\theta)}{d\theta_{i}} &  =x_{n,i}^{\prime}((\xi_{n,m}%
)^{+})(t_{n,m}^{1}-\xi_{n,m})+x_{n,i}^{\prime}((t_{n,m}^{J_{n,m}})^{+}%
)\cdot\nonumber\\
&  (\eta_{n,m}-t_{n,m}^{J_{n,m}})+\sum\limits_{j=2}^{J_{n,m}}x_{n,i}^{\prime
}((t_{n,m}^{j})^{+})(t_{n,m}^{j}-t_{n,m}^{j-1})\label{IPAestimator}
\end{align}
Clearly, $x_{n,i}^{\prime}$ at each event time is determined by (\ref{jumps})
which in turn depends on the event type at $t_{n,m}^{j}$, $j=1,...,J_{n,m}$
and is given by the corresponding expression in (\ref{E_event}),
(\ref{type2di2}), (\ref{type3}), and (\ref{4a})-(\ref{type4bc}). The
associated event time derivatives in these expressions are provided
recursively through (\ref{2a}), (\ref{2b})-(\ref{g2r_i2}) and (\ref{3a})-(\ref{r2g_i2}). Note that $dL_{n,m}(\theta)/d\theta_{i}$ can
be computed using an on-line algorithm that updates $\tau_{k,i}^{\prime}$ and
$x_{n,i}^{\prime}$ after every observed event. More importantly, this IPA
derivative depends on: $(i)$ the number of events in each NEP $J_{n,m}$,
$(ii)$ the number of $G2R_{n}$ events and $R2G_{n}$ events, $(iii)$ the event
times $\xi_{n,m}$, $\eta_{n,m}$ and $t_{n,m}^{j}$, and $(iv)$ the arrival and
departure rates $\alpha_{n}(\tau_{k})$, $\beta_{n}(\tau_{k})$ at an event time
\emph{only}. The quantities in $(i)-(iii)$ are easily observed through counters
and timers. The rates in $(iv)$ may be obtained through simple rate estimators,
emphasizing that they are \emph{only} needed at a finite number of observed
event times. As already pointed out, the detailed nature of
$\alpha_{n}(t)$, $\beta_{n}(t)$ is not required in (\ref{IPAestimator}).

\section{Simulation Results}

We describe how the IPA estimator derived for the SFM can be used to improve
performance and ultimately determine optimal light cycles for an intersection modeled as a Discrete Event System (DES). We apply the IPA estimator using actual
data from an observed sample path of this DES (in this case, by simulating it
as a pure DES).

We assume vehicles arrive according to a Poisson process with rate
$\bar{\alpha}_{n}$, $n=1,2$ (as already emphasized, our results are
independent of this distribution.). We also assume vehicles depart at a fix rate
$\beta_{n}$ if the road is not empty. We constrain $\theta_{i,1},i=1,2$, to
take values in $[\theta_{\min,1},\theta_{\max,1}]$, and constrain
$\theta_{i,2}$, to take values in $[\theta_{i,1},\theta_{\max,2}]$.

For the simulated DES model, we use a brute-force (BF) method to find an
optimal $\theta_{BF}^{\ast}$: we discretize all real values of $\theta_{i}$
and for combinations of  $\theta_{i},i=1,...,4$ (based on our previous definition, $\theta_{1}\equiv\theta_{1,1},\theta_{2}\equiv\theta_{1,2},\theta_{3}\equiv\theta_{2,1},\theta_{4}\equiv\theta_{2,2}$), we run $10$ sample paths to
obtain the average total cost. The value of $\theta_{BF}^{\ast}$ is the one
generating the least average cost, to be compared to $\theta_{IPA}^{\ast}$,
the IPA-based method. In our simulations, we estimate $\alpha_{n}(\xi_{n,m})$
through ${N_{a}}/{t_{w}}$ by counting vehicle arrivals $N_{a}$ over a time
window $t_{w}$ before or after $\xi_{n,m}$; $\beta_{n}(\tau_{k})$ is similarly estimated.

\begin{figure}[tb]
\centering
\includegraphics[scale = 0.4]{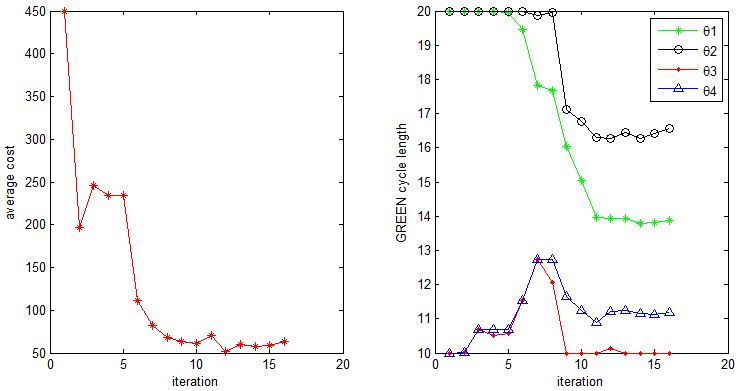}\caption{Sample cost ($J$)
and parameter $(\theta)$ trajectories}%
\label{CDC13Simu1}%
\end{figure}

In all our simulations, we set $\theta_{\min,1}=10sec$, $\theta_{\max
,1}=20sec$, $\theta_{\max,2}=40sec$, $\beta_{1}=\beta_{2}=1$ and $T=2000sec$.
We also set the weight $w_{i}=1$ if $x_{i}<S_{i}$, and $w_{i}=10$ if
$x_{i}\geq S_{i},i=1,2$, which indicates there is more cost if the queue
content exceeds the threshold. In Fig. \ref{CDC13Simu1}, we show sample
trajectories of $J$ and $\theta$ where we set $1/\alpha=[1.9,3]$ and
$S=[8,8]$. As we can see, the gradient-based algorithm converges quickly with
an optimal cost $J_{IPA}^{\ast}=51.68<$ $J_{BF}^{\ast}=62.06$ obtained by the
BF method. Extensive additional comparison results under different traffic
intensities (denoted by $\alpha$) are provided in Table \ref{CDC13Table1}.
Generally, the IPA method gives similar performance with the BF method, which,
however, becomes impractical when there are more controllable parameters, or
when the ranges of the parameters are large. 

\begin{table}[tbh]
\caption{IPA vs BF method with different $\alpha$}%
\label{CDC13Table1}
\begin{center}%
\begin{tabular}
[c]{|c|c|c||c|c|}\hline
\multirow{2}{*}{$1/\bar{\alpha}$} & \multicolumn{2}{|c||}{BF} &
\multicolumn{2}{|c|}{IPA}\\\cline{2-5}
& $\theta^{*}$ & $J^{*}$ & $\theta^{*}$ & $J^{*}$\\\hline
{[2.2,2.7]} & [10,20,10,14] & 12.7 & [10,15.1,11.3,11.3] & 12.4\\\hline
{[2,3]} & [11,30,10,14] & 12.3 & [10.2,19.3,10.1,16.3] & 10.9\\\hline
{[1.9,3]} & [10,30,10,18] & 16.4 & [18.9,25.6,13.3,17.0] & 16.3\\\hline
{[1.8,3]} & [14,30,10,20] & 17.6 & [10.1,20.0,10.1,12.1] & 15.7\\\hline
{[1.7,3]} & [12,26,10,12] & 24.6 & [10.1,20.1,10.6,11.9] & 25.9\\\hline
\end{tabular}
\end{center}
\end{table}

\begin{figure}[tbh]
\centering
\includegraphics[scale = 0.4]{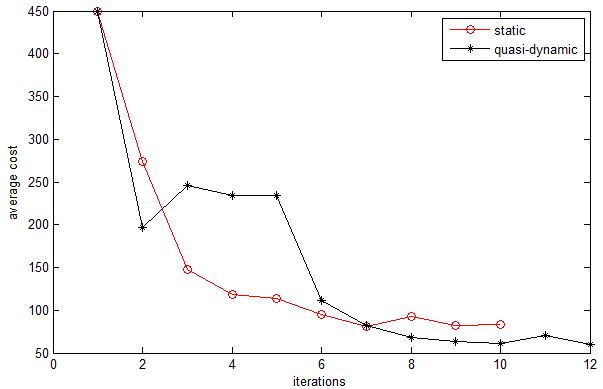}\caption{Sample cost
$(J)$ trajectories using two static and quasi-dynamic control}%
\label{CDC13Simu2}%
\end{figure}

\begin{figure}[tbh]
\centering
\includegraphics[scale = 0.45]{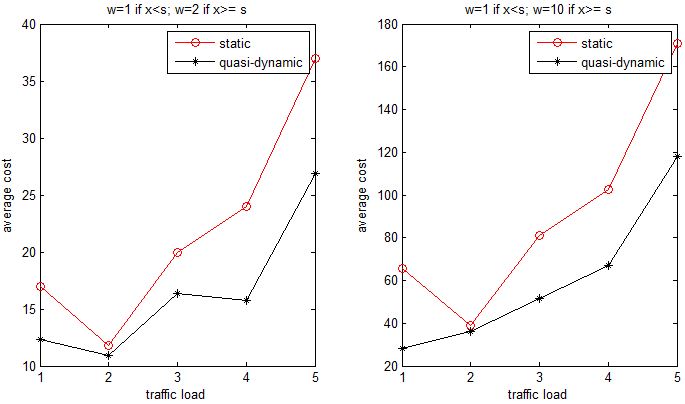} \caption{Optimal cost
comparisons under different traffic loads}%
\label{CDC13Simu3}%
\end{figure}

Of greater interest is a comparison of the quasi-dynamic control to the one in
\cite{GengCDC12}, which uses a static controller based on $\theta^{\ast}$ in
between two adjustment points. We use the same settings for the two models,
where we constrain $\theta\in\lbrack10,40]$ in \cite{GengCDC12}. In Fig.
\ref{CDC13Simu2} we show the cost trajectories of the two methods using the
settings as in Fig. \ref{CDC13Simu1} and the same initial $\theta_{0}%
=[20,10]$. The quasi-dynamic IPA-based method converges to a lower cost. More
comparison results are shown in Fig. \ref{CDC13Simu3} where the $x$-axis
denotes the five traffic intensities in Table \ref{CDC13Table1}. As we can
see, using quasi-dynamic control generally results in better performance than
using static control, as expected. 

\begin{figure}[tbh]
\centering
\includegraphics[scale = 0.45]{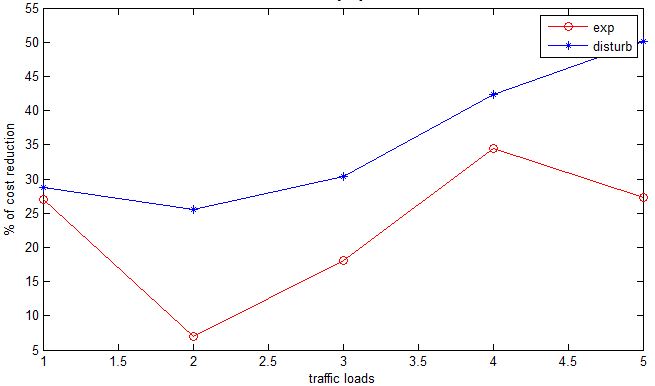} \caption{Cost reduction
comparisons under different traffic loads}%
\label{CDC13Simu4}%
\end{figure}

In all simulations above, we use exponentially distributed interarrival time for
the incoming traffic flow. Thus the traffic generally does not significantly
vary due to the relatively low variance of the Poisson process. We expect to
see more drastic improvement using our method under highly variable traffic.
To justify this, we randomly add a burst of traffic into road 1 and observe
the resulting \textquotedblleft adaptivity\textquotedblright\ of our method.
In Fig. \ref{CDC13Simu4} we compare the cost reduction under purely
exponential traffic (exp) with the one under \textquotedblleft
disturbed\textquotedblright\ traffic (disturb). The $y$-axis denotes the
percentage of cost reduction: $\frac{J_{fixed}^{\ast}-J_{IPA}^{\ast}%
}{J_{fixed}^{\ast}}\ast100$, where $J_{fixed}^{\ast}$ is the optimal value
obtained using the IPA method in \cite{GengCDC12}. As we can see,
quasi-dynamic control exhibits higher cost reduction under highly variable traffic.

\section{Conclusions and Future work}

We have developed a SFM for a single intersection traffic light control
problem, based on which we derive an IPA gradient estimator of a cost metric
with respect to the controllable green and red cycle lengths. The estimator is
used to iteratively adjust light cycle lengths to improve performance and,
under proper conditions, obtain optimal values which adapt to changing traffic
conditions. In contrast to prior work \cite{GengCDC12}, we apply quasi-dynamic
cycle length control between adjustment points, using partial state
information defined by detecting whether a vehicle backlog is above or below a
certain threshold, without the need to observe an exact vehicle count.
Numerical results show that the IPA method leads to better performance than
that obtained through repetitive \textquotedblleft
brute-force\textquotedblright\ simulation, and better than using static cycle
control. Future work will focus on using IPA to solve the TLC problem for an
intersection with more complicated traffic (e.g., left-turn and right-turn),
modeling accelerating traffic following a GREEN light and extending our method
to solving the TLC problem over multiple junctions. Moreover,
we will explore the use of IPA in controlling the queue content thresholds in
addition to the light cycle lengths.

\bibliographystyle{IEEEtran}
\bibliography{yf}

\end{document}